\documentstyle[preprint,revtex,eqsecnum]{aps}
\narrowtext
\def\ybcox{YBa$_2$Cu$_3$O$_{6+x}$}
\def\lnco{La$_{2-x}$Nd$_x$CuO$_4$}
\tightenlines
\begin{document}
%
\preprint{ETH-TH/92-31}
\begin{title}
Theory of Anisotropic Superexchange in Insulating Cuprates
\end{title}
\author{N.\ E.\ Bonesteel}
\begin{instit}
Theoretische Physik, Eidgen\"ossische Technische
Hochschule-H\"onggerberg,\\ CH-8093 Z\"urich, Switzerland
\end{instit}
\begin{abstract}
Spin-orbit corrections to superexchange are calculated using the
method of Moriya [T.\ Moriya, {\it Phys.\ Rev.}~{\bf 120} 91, (1960)]
for two of the insulating parent compounds of the cuprate
superconductors: (1) La$_{2-x}$Nd$_x$CuO$_4$ where the CuO$_6$
octahedra forming each Cu-O layer are tilted in staggered fashion
about an axis which depends on $x$ and temperature; and (2)
YBa$_2$Cu$_3$O$_{6+x}$ ($x \alt 0.4$) where the Cu-O layers form
CuO$_2$-Y-CuO$_2$ bilayers in which the in-plane O$^{2-}$ ions are
displaced uniformly towards the Y$^{3+}$ layer.  For (1) a simple
formula is derived for the weak ferromagnetic moment in each Cu-O
layer as a function of the tilting axis and magnitude.  For (2) it is
shown that the anisotropic corrections to superexchange are different
from what has previously been assumed.  For the correct spin
Hamiltonian a classical N\'eel state in which the Cu spins are lying
in the plane is unstable in a single Cu-O layer, but when a bilayer is
considered there is a critical value of the interlayer exchange
coupling which stabilizes this state.  For both cases (1) and (2)
spin-wave spectra are calculated and shown to compare favorably with
experiment.
\end{abstract}
\pacs{      }
\narrowtext
%
\section{Introduction}

Spin-orbit (SO) coupling causes electron spins to precess as they move
through the electric field of a crystal lattice.  Within the
tight-binding approximation this precession appears as a small spin
rotation which occurs whenever an electron tunnels between two Wannier
orbitals.  As first shown by Moriya \cite{moriya} this rotation can
have important consequences in antiferromagnetic (AFM) Mott
insulators; when it is included in Anderson's calculation \cite{pwa}
of superexchange then anisotropic corrections to the otherwise
isotropic effective spin Hamiltonian are generated.  These
corrections, known as Dzyaloshinski-Moriya (DM) interactions
\cite{moriya,dizzie,note1}, lift any ground state degeneracy
associated with rotational invariance in spin space and are
responsible for such effects as weak ferromagnetism and the existence
of spin-wave anisotropy gaps.

The subject of this paper is the DM interactions which exist in the
distorted Cu-O layers of the insulating AFM parent phases of the
cuprate superconductors.  We will be concerned both with the
microscopic origin of these interactions and their physical
consequences.  Probably the best known example of the effects of DM
interactions in the cuprates occurs in La$_2$CuO$_4$ \cite{wf}.  In
this material a structural phase transition occurs from a high
temperature tetragonal (HTT) phase (space group I4/{\it mmm}) to a low
temperature orthorhombic (LTO) phase (space group {\it Bmab}).  In the
LTO phase the CuO$_6$ octahedra forming each Cu-O layer tilt in a
staggered pattern about the $\langle 1\overline{1}0 \rangle$ axis and
this distortion results in DM interactions which induce a weak
ferromagnetic moment in each layer \cite{wf}.  Another example of the
effects of DM interactions in the cuprates is the easy-plane
anisotropy which has been observed in the spin-wave spectrum of the
AFM insulating phase of YBa$_2$Cu$_3$O$_{6+x}$ ($x \alt 0.4$)
\cite{rm}.  In this case the relevant structural feature is that the
Cu-O layers form CuO$_2$-Y-CuO$_2$ bilayers in which the negatively
charged in-plane O$^{2-}$ ions are uniformly buckled towards the
positively charged Y$^{3+}$ layer.

One motivation for the present work is that in a recent paper
\cite{spirals} a `one-band' description of SO coupling in both
insulating and doped La$_2$CuO$_4$ in the presence of various tilting
distortions was studied using the Hamiltonian
\begin{equation}
H_{\rm flux} = -t\sum_{{\scriptstyle
\langle ij\rangle}\atop{\scriptstyle
\alpha}}\left\{e^{i \phi_{ij}\alpha} c^\dagger_{i\alpha}
c^{\phantom{\dagger}}_{j\alpha}+h.c.\right\}+U\sum_{i} n_{i\uparrow}
n_{i\downarrow}.\label{pureflux}
\end{equation}
Here $c^\dagger_{i\alpha}$ creates an electron with spin $\alpha$ at
site $i$, $n_{i\alpha}$ is the corresponding number operator,
$\phi_{i,i+{\hat{\bf x}}}\simeq (-1)^{(x_i + y_i)} 0.05\theta$,
$\phi_{i,i+{\hat{\bf y}}} = -\phi_{i,i+{\hat{\bf x}}}$ where $\theta$
is the octahedral tilt angle, and in the exponent $\alpha= +(-)1/2$
for up (down) spins.  The $x$ and $y$ components of site $i$ are
denoted $x_i$ and $y_i$, and $\hat{\bf x}$ and $\hat{\bf y}$ are unit
vectors.  Hamiltonian (\ref{pureflux}) describes a correlated
tight-binding band of electrons which move in a background of
staggered flux where up and down spin electrons have opposite charge.
This flux is simply the Berry's phase associated with spin precession
about the $z$-axis in spin space.  In \cite{spirals} the effective
Hamiltonian describing the large $U/t$ limit of (\ref{pureflux}) was
derived and at half-filling the classical spin ground state was found
to have no weak ferromagnetic moment.  It is shown here that this
result is not in conflict with the experimental observation of such a
moment in the LTO phase of La$_2$CuO$_4$.  Hamiltonian
(\ref{pureflux}) is a correct description of electrons in a tilted
Cu-O layer when the system is viewed in the appropriate {\it
site-dependent coordinate system in spin space}.  When local rotations
are performed to transform the system back to the physical spin-space
coordinate system a weak ferromagnetic moment which agrees with
experiment appears.

A similar approach to weak ferromagnetism has been discussed recently
by Shekhtman, Entin-Wohlman and Aharony (SEA) \cite{sea} who also
studied the DM interactions induced by tilting distortions in
La$_2$CuO$_4$.  SEA were able to show that, quite generally, the DM
interactions present on a single Cu-O-Cu bond are isotropic when the
bond is viewed in the appropriate local coordinate system in spin
space.  This observation led them to the interesting conclusion that
any physical anisotropy ({\it i.e.}, anisotropy which cannot be
`gauged away' by local rotations) must arise from the {\it
frustration} of these bonds.  By applying this idea to the LTO phase
of La$_2$CuO$_4$ SEA were able to successfully account for the
observed weak ferromagnetism in this material.  In this paper we
calculate the DM interactions which occur in the presence of tilting
distortions in which the CuO$_6$ octahedra can tilt about any axis,
not just $\langle 1\overline{1}0 \rangle$.  Such general tilting
distortions may have physical relevance because they describe at least
the average structure of La$_{2-x}$Nd$_x$CuO$_4$ when $x \alt 0.5$
\cite{pccn1,pccn2}.  The main result of our analysis is that
regardless of the octahedral tilt axis the effective Hamiltonian
describing a single layer can always be transformed into
(\ref{pureflux}), to lowest order in $\theta$.  However, the
coordinate system in spin space in which (\ref{pureflux}) holds
changes as the tilt axis changes so that the ratio of the weak
ferromagnetic moment to the size of the tilting distortion depends on
the tilt axis in a simple way which we derive in Section IV.  These
results provide a potential experimental test of the `anisotropy
through frustration' idea of SEA \cite{sea} and justifies the use of
(\ref{pureflux}) in \cite{spirals}.

A second motivation for this work is the puzzling observation by
Coffey, Rice and Zhang \cite{crz} that the DM interactions in a single
buckled Cu-O layer of YBa$_2$Cu$_3$O$_{6+x}$ tend to stabilize an
incommensurate spiral spin configuration, while neutron scattering
experiments see no sign of such a spiral \cite{rm}.  It is shown here
that if an effective spin Hamiltonian for a single Cu-O layer in
YBa$_2$Cu$_3$O$_{6+x}$ is derived using the same method as for
La$_{2-x}$Nd$_x$CuO$_4$ then a classical N\'eel state with spins lying
in the $xy$ plane is indeed unstable; complex frequencies
corresponding to exponentially growing unstable modes appear in the
classical linearized spin-wave spectrum.  However, when instead of a
{\it single} Cu-O layer a CuO$_2$-Y-CuO$_2$ {\it bilayer} is
considered these complex frequencies disappear for a critical value of
the interlayer coupling $J_{12}^c \simeq 1\times10^{-3} J$, well below
the lower limit on $J_{12}$ set by experiment \cite{rm}, and the
N\'eel state becomes stable.  This stabilization occurs because the DM
interactions in the upper and lower planes favor spirals with opposite
senses (a consequence of the inversion symmetry of the
YBa$_2$Cu$_3$O$_{6+x}$ unit cell) and this spiraling is frustrated by
the interlayer coupling.  We also find that when $J_{12}$ is large
enough not only does the N\'eel state become stable, but the spin-wave
spectrum shows an in-plane gapless mode and an out-of-plane gapped
mode in agreement with experiment \cite{rm}.  This result shows that
the `easy-plane' anisotropy in YBa$_2$Cu$_3$O$_{6+x}$ in fact arises
from frustration of DM interactions in accordance with the general
principle of SEA
\cite{sea}.

This paper is organized as follows.  In Section II the structural and
magnetic properties of La$_{2-x}$Nd$_x$CuO$_4$ and
YBa$_2$Cu$_3$O$_{6+x}$ which are relevant for this paper are
discussed.  The SO modification of superexchange due to the structural
distortions in these materials is calculated in Section III and the
resulting classical ground state and spin-wave excitation spectra are
presented in Section IV and V, respectively.  Finally, Section VI
summarizes the conclusions of the paper.

\section{Relevant Experimental Facts}

\subsection{La$_{2-x}$Nd$_x$CuO$_4$}
The La$_{2-x}$Nd$_x$CuO$_4$ system shows a rich structural phase
diagram as a function of $x$ and temperature.  As mentioned above when
$x = 0$ the material undergoes a phase transition from the HTT phase
to the LTO phase.  It is also known that when enough Nd is doped into
the system ($x > 0.5$) the material crystallizes into the T$^\prime$
structure of pure Nd$_2$CuO$_4$ \cite{bts}.  However, for smaller Nd
concentration $(x \alt 0.4)$ there is still an HTT $\rightarrow$ LTO
transition, and as the temperature is lowered further there is a
second transition into a structural phase with space group {\it Pccn}
\cite{pccn1,pccn2}.  In both the LTO and {\it Pccn} phases the CuO$_6$
octahedra forming each Cu-O layer tilt in a staggered fashion through
an angle $\theta (\simeq 0.1)$ about first the $(\cos\chi,\sin\chi,0)$
and then the $(\sin\chi,\cos\chi,0)$ axis in successive Cu-O layers
where $\chi = \pi/4$ in the LTO phase and $0<\chi<\pi/4$ in the {\it
Pccn} phase.  The case $\chi = 0$ corresponds to the low temperature
tetragonal (LTT) phase (space group P4$_2$/{\it ncm}) which occurs,
for example, in the doped material La$_{1.88}$Ba$_{0.12}$CuO$_4$
\cite{ltt}.

Neutron scattering measurements of the spin structure factor of
insulating La$_2$CuO$_4$ and subsequent theoretical analysis have
shown fairly conclusively that the spin degrees of freedom in this
material are well described by an AFM Heisenberg model with exchange
coupling $J \simeq$ 130 meV \cite{chakra}.  Although this model is
adequate for describing most properties of La$_2$CuO$_4$ slight
deviations from perfect isotropy have been observed experimentally.
In particular: (i) Thio {\it et al.} and Cheong {\it al.}
\cite{wf} observed a first order spin-flop transition as a function of
applied magnetic field perpendicular to the Cu-O planes -- this was
interpreted as being due to the existence of weak ferromagnetic
moments in each layer; and (ii) both neutron scattering
\cite{swg1} and AFM resonance measurements \cite{swg2} have shown that
the zone-center spin-waves in La$_2$CuO$_4$ are gapped with an
in-plane gap of $\sim$ 1.0 meV and an out-of-plane gap of $\sim$ 2.5
meV.  It was immediately realized by the groups which performed these
measurements that these effects were manifestations of DM
interactions.  These interactions have since been calculated
microscopically \cite{crz,sea,kom}.  The resulting spin Hamiltonian
has an Ising-like anisotropy, which is responsible for the zone-center
spin-wave gaps, and a ground state in which the spins lie nearly along
the orthorhombic $c$ axis except for a slight cant out of the Cu-O
plane which gives each layer a weak ferromagnetic moment.  The canting
angle is roughly $\Theta_{wf}\sim0.005$ and so the weak ferromagnetic
moment is $\sim 0.003 \mu_{B}$ per Cu site.

\subsection{YBa$_2$Cu$_3$O$_{6+x}$}
For $x \alt 0.4$ the YBa$_2$Cu$_3$O$_{6+x}$ system is tetragonal and
AFM with the added oxygens going into the chains presumably at random.
When $x \agt 0.4$ the oxygen ions partially order and this leads to a
tetragonal $\rightarrow$ orthorhombic transition.  At the same doping
the planes become metallic and the material becomes a superconductor.
Here we are primarily concerned with the tetragonal insulating AFM
phase.  The key structural feature in this material is that the Cu-O
layers are not equally spaced as they are in the La system but instead
form CuO$_2$-Y-CuO$_2$ bilayers.  The planes forming these bilayers
are buckled with the in-plane negatively charged O$^{2-}$ ions
displaced uniformly towards the positively charged Y$^{3+}$ layer.
The size of the oxygen dispacement out of the plane depends only
weakly on $x$ and is roughly 0.22 \AA\
\cite{swst} and so the CuO bonds make an angle of roughly
$\theta\simeq0.1$ with the Cu-O plane \cite{swst}.  Thus the magnitude
of the distortion is approximately the same as in La$_2$CuO$_4$.

Previously the magnetic structure of YBa$_2$Cu$_3$O$_{6+x}$ has been
modeled assuming each bilayer can be described by a Hamiltonian of
the form
\cite{alloul}
\begin{equation}
H = \sum_{a=1,2}\sum_{ij}\left\{ J_{z}S^{z}_{a,i}S^{z}_{a,j} +
J_{xy}(S^{x}_{a,i}S^{x}_{a,j}
+S^{y}_{a,i}S^{y}_{a,j})\right\}+J_{12}\sum_{i}\vec S_{1,i}\cdot\vec
S_{2,i}\label{epb}
\end{equation}
The spin-wave spectrum of (\ref{epb}) has four branches: a gapless
in-plane mode; a gapped out-of-plane mode with a gap of
$4S\sqrt{2(J_{xy} - J) J}$; and two high-energy branches with gaps of
$4\sqrt{2}S\sqrt{JJ_{12}}$.  Neutron scattering has shown the
existence of a gapless in-plane mode with spin-wave velocity $\sim$1.0
eV-\AA\ ($J \simeq 150$ meV) and a gapped out-of-plane mode with gap
$\sim 4$ meV \cite{rm} and so the low-energy spectrum of (\ref{epb})
agrees with the experiment.  At the same time the high-energy modes
have not been observed for energy transfers up to $50$ meV putting a
lower limit on the interlayer coupling of $J_{12}\agt 0.01 J$
\cite{rm,tranq}.  The microscopic basis of (\ref{epb}) is not as firm
as that of the corresponding spin Hamiltonian for La$_2$CuO$_4$ and in
Section V we will show that, in fact, when the same methods which have
been used successfully to describe La$_2$CuO$_4$ are applied to
YBa$_2$Cu$_3$O$_{6+x}$ the resulting spin Hamiltonian is different.
Nonetheless, it is possible to reproduce the experimentally observed
spin-wave spectrum using the new Hamiltonian.

\section{Spin-Orbit Corrections to\\ Superexchange}
Superexchange occurs when two magnetic ions interact through their
mutual overlap with an intermediate diamagnetic ion \cite{pwa}.
Recently SEA \cite{sea} have used Moriya's method \cite{moriya} to
derive a fairly general expression for the anisotropic corrections to
superexchange due to SO coupling on a single such bond for spin-1/2.
Their result, which corrects some omissions in an earlier calculation
\cite{crz} as  well as a numerical  error  in Moriya's original paper,
has the interesting property that it can be related to an isotropic
interaction by a unitary transformation \cite{sea}.  A similar
expression for the anisotropic superexchange on a single bond was
implicit in \cite{spirals} and in this section its derivation is
sketched to show that in fact the expressions in \cite{spirals} and
\cite{sea} are the same and to establish notation for the rest of the
paper.

In the presence of SO coupling a single Cu-O-Cu bond is described by
the Hamiltonian
\cite{moriya,crz,sea,kom}
\widetext
\begin{eqnarray}
H_{\rm CuOCu} =&&
\sum_{\alpha\beta}\Bigl\{d^\dagger_{1\alpha}(t_{pd}\delta_{\alpha\beta}
+i\vec\lambda_{1}\cdot\vec\sigma_{\alpha\beta})
p^{\phantom{\dagger}}_{\beta}\nonumber
+d^\dagger_{2\alpha}(t_{pd}\delta_{\alpha\beta}
+i\vec\lambda_{2}\cdot\vec\sigma_{\alpha\beta})
d^{\phantom{\dagger}}_{\beta}+h.c.\Bigr\}
\nonumber \\
&&+U_{dd}( n^{\phantom{\dagger}}_{d1\uparrow}
n^{\phantom{\dagger}}_{d1\downarrow}
+n^{\phantom{\dagger}}_{d2\uparrow}n^{\phantom{\dagger}}_{d2\downarrow})
+\Delta_{dp}\sum_\alpha
p^\dagger_{\alpha}p^{\phantom{\dagger}}_{\alpha}\label{se}
\end{eqnarray}
\narrowtext\noindent
The vacuum state for (\ref{se}) is one in which the Cu $3d$ and O $2p$
shells are full.  The operators $d^\dagger_{i\alpha}$ and
$p^\dagger_{\alpha}$ then create holes with spin $\alpha$ at Cu site
$i$ and the oxygen $\sigma$ orbital, respectively.  The (hole) energy
splitting between Cu and O sites is $\Delta_{pd}$, the on-site
correlation on the Cu site is $U_{dd}$, and $t_{pd}$, $\vec\lambda_1$
and $\vec\lambda_2$ are hopping integrals, the latter two arising from
SO coupling \cite{sea,moriya}.

The problem of deriving the effective  spin Hamiltonian for (\ref{se})
simplifies upon applying the unitary transformation
\begin{eqnarray}
{d^\prime}^\dagger_{1\beta} &=& \sum_\alpha
\left[\exp\left(i\tan^{-1}({|\vec\lambda_1|\over{t_{pd}}})
{\vec\lambda_{1}\cdot\vec\sigma\over{2|\vec\lambda_1
|}}\right)\right]_{\alpha\beta}d^\dagger_{1\alpha}
\nonumber\\
{d^\prime}^\dagger_{2\beta} &=&
\sum_\alpha
\left[\exp\left(i\tan^{-1}({|\vec\lambda_2|\over{t_{pd}}})
{\vec\lambda_{2}\cdot\vec\sigma\over{2|\vec\lambda_2
|}}\right)\right]_{\alpha\beta} d^\dagger_{2\alpha}.\label{set}
\end{eqnarray}
When (\ref{se}) is expressed in terms of the primed operators the
result is
\begin{equation}
H_{\rm CuOCu} = \sum_\alpha \left\{
\tilde t_{pd} {d^\prime}^\dagger_{1\alpha} p^{\phantom{\dagger}}_{\alpha}
+\tilde t_{pd}
{d^\prime}^\dagger_{2\alpha}p^{\phantom{\dagger}}_{\alpha}
+h.c.\right\}
+U( n^{\phantom{\dagger}}_{1\uparrow}
n^{\phantom{\dagger}}_{1\downarrow} +n^{\phantom{\dagger}}_{2\uparrow}
n^{\phantom{\dagger}}_{2\downarrow}) +\Delta\sum_\alpha
p^\dagger_\alpha p^{\phantom{\dagger}}_\alpha\label{stransse}
\end{equation}
where $\tilde t_{pd} = (t_{pd}^2+\lambda^2)^{1/2}$, (we will consider
only the case where $|\vec\lambda_1| = |\vec\lambda_2| = \lambda$).
The transformation (\ref{set}) absorbs the spin precession induced by
SO coupling into a redefinition of the local coordinate system in spin
space.  Such a transformation is possible because the bond is
essentially one dimensional; {\it i.e.}, there are no closed loops
around which an electron can hop and acquire a finite spin precession
which cannot be transformed away.

The effective spin Hamiltonian resulting from (\ref{se}) in the limit
$\tilde t_{pd} \gg U_{dd},\Delta_{pd}$ can now be found using standard
methods \cite{emery}.  The result is $H_{\rm Bond} = J\vec
S^\prime_1\cdot \vec S^\prime_2$, with
\begin{equation}
J={4{\tilde t}_{dp}^4\over\Delta_{dp}^2}
\left\{{{1\over\Delta_{dp}}+{1\over U_{dd}}}\right\}
\label{jemery}
\end{equation}
and $\vec S^\prime_1 = (1/2){d^\prime}^\dagger_{1\alpha} \vec
\sigma_{\alpha\beta} {d^\prime}^{\phantom{\dagger}}_{1\beta}
+ O(t/[\Delta_{pd},U])$.  When the unitary transformation (\ref{set})
is undone the final result is
\begin{equation}
H_{\rm Bond} = J\Bigl(S^z_1 S^z_2
+\cos\phi (S^x_1 S^x_2 + S^y_1 S^y_2)
+\sin\phi (S^x_1 S^y_2 - S^y_1
S^x_2)\Bigr)\label{dmbond}
\end{equation}
where $\phi\simeq |\vec\lambda_1-\vec\lambda_2|/t_{pd}$ is the angle
through which an electron spin precesses when it hops from site $1$
through the intermediate orbital to site $2$, and where the $z$-axis
in spin space has been chosen to be parallel to the precession axis
$\vec\lambda_1-\vec\lambda_2$.

After some algebra it is possible to show that (\ref{dmbond}) is
equivalent to the result obtained by SEA \cite{sea}. Hamiltonian
(\ref{dmbond}) was also derived in precisely the form given above in
\cite{spirals} but starting from a `one-band' description in which the
oxygen ions were not included explicitly.  The equivalence between the
one-band and three-band pictures at half filling is easy to
understand.  If the Hamiltonian for a single bond is given by
\begin{equation}
H_{\rm 1-band} =
\sum_\alpha\left\{c^\dagger_{1\alpha}(-t\delta_{\alpha\beta}+
i\vec\lambda_{12}\cdot
\vec\sigma_{\alpha\beta})c^\dagger_{2\alpha}+h.c.\right\}
+U\left( n_{1\uparrow}n_{1\downarrow}
+n_{2\uparrow}n_{2\downarrow}\right),
\label{oneband}
\end{equation}
and if for a given $t$ we choose $U$ so that $4(t^2 + \lambda_{12}^2)/
U = J$ and $\vec\lambda_{12} \simeq t/t_{pd} (\vec\lambda_1 -
\vec\lambda_2)$, then similar arguments to those given above yield
(\ref{dmbond}) when $U\gg t$.  In what follows we will adopt this
`one-band' approach and describe a given Cu-O layer with a Hamiltonian
of the form
\begin{equation}
H = \sum_{\scriptstyle{\langle ij
\rangle}\atop{\scriptstyle\alpha\beta}}
\left\{c^\dagger_{i\alpha} \left(-t\delta_{\alpha\beta}
+ i \vec\lambda_{ij} \cdot \vec\sigma_{\alpha\beta}\right)
c^{\phantom{\dagger}}_{\beta}+{\rm h.c.} \right\}
+U\sum_{i}
n^{\phantom{\dagger}}_{i\uparrow}n^{\phantom{\dagger}}_{i\downarrow}.\label{tb}
\end{equation}
At half filling there exists an entire class of models with different
$t$, $U$, and $\vec\lambda_{ij}$ values which yield the same effective
spin Hamiltonian in the large $U/t$ limit.  Away from half filling
this is no longer the case and it is necessary to perform a mapping
from the full three-band model to an effective one-band model in order
to find the appropriate parameters \cite{spirals}.  In what follows we
are only concerned with the half-filled case and so for simplicity we
take $t = t_{pd} \simeq 1.3$ eV \cite{hyb}.

It remains to compute the $\vec\lambda_{ij}$ vectors.  At half-filling
each Cu ion in a Cu-O layer has one hole in its 3$d$ shell which, in
the absence of SO coupling, occupies the $d_{x^2-y^2}$ orbital.  If on
each Cu site a SO interaction
\begin{equation}
H_{SO} = \beta\sum_i\vec L_i\cdot\vec S_i
\end{equation}
is included where $\vec L_i$ and $\vec S_i$ are the orbital angular
momentum and the hole spin at site $i$, respectively, and $\beta\simeq
0.1$ eV for Cu \cite{slater}, then higher crystal field levels --
$d_{xy}$, $d_{xz}$, and $d_{yz}$ -- are mixed into the lowest lying
$d_{x^2-y^2}$ state.  These admixtures then modify the hopping
integrals and give rise to the $\vec\lambda$ terms in (\ref{se}).

Moriya \cite{moriya} derived an expression in second-order
perturbation theory for the $\vec\lambda_{ij}$ vectors in (\ref{tb})
which for our purposes reads:
\begin{equation}
\vec\lambda_{ij}\simeq\vec\lambda_i-\vec\lambda_j
\simeq {i\beta \over 2} \sum_m \Bigl\{ {\langle
m,i|{\vec L}_i|0,i\rangle
\over{\epsilon_m-\epsilon_0}}t_{ij}(m,\sigma) -{\langle m,j|{\vec
L}_j| 0,j\rangle^* \over {\epsilon_m-\epsilon_0}}t_{ji}(m,\sigma)
\Bigr\}\label{moriya}
\end{equation}
where $|m,i\rangle$ is a crystal field split level in the absence of
SO coupling labeled by $m$ at site $i$ with $m=0$ corresponding to
$d_{x^2-y^2}$.  The energy of the $m$th level is $\epsilon_m$, and
$t_{ij}(m,\sigma)$ is the hopping matrix element in the absence of SO
coupling between the Cu orbital $m$ at site $i$ and the oxygen
$\sigma$ orbital between sites $i$ and $j$.  The relevant matrix
elements of $\vec L$ are
\begin{eqnarray}
\langle {x^2-y^2},j | \vec L_j |xz,j\rangle &=& i{\hat{\bf y}},\nonumber\\
\langle {x^2-y^2},j | \vec L_j |yz,j\rangle &=& -i{\bf {\hat x}},\label{lme}
\end{eqnarray}
and so to evaluate (\ref{moriya}) only the matrix elements
$t_{ij}(m,\sigma)$ need to be determined.  This is done below for
La$_{2-x}$Nd$_x$CuO$_4$ and YBa$_2$Cu$_3$O$_{6+x}$.

\subsection{La$_{2-x}$Nd$_x$CuO$_4$}
In the LTO and {\it Pccn} phases of La$_{2-x}$Nd$_x$CuO$_4$ the
average structure is one in which the CuO$_6$ octahedra forming each
Cu-O layer are rotated through an angle $(-1)^{(x_i+y_i)}
\theta$ about the axis $(\cos\chi,\sin\chi,0)$.  In general there are
then two CuO bond angles: $\pm \theta\sin\chi$ and $\pm\theta\cos\chi$
for bonds pointing in the $x$ and $y$ directions, respectively.
Figure (\ref{fig1}a) shows the Cu $xz$ orbitals and oxygen $\sigma$
orbital in a typical bond in La$_{2-x}$Nd$_x$CuO$_4$.  For this bond
the hopping from the rotated $xz$ orbitals to the $\sigma$ orbital is,
for small $\theta$,
\begin{eqnarray}
t_{i,i+{\hat{\bf x}}}(xz,\sigma) &\simeq & (-1)^{x_i +y_i}
V_{pd\pi}\theta\sin\chi,\nonumber \\
t_{i,i+{\hat{\bf x}}}(yz,\sigma) &\simeq & 0\label{t1214}
\end{eqnarray}
and similarly
\begin{eqnarray}
t_{i,i+{\hat{\bf y}}}(xz,\sigma) &\simeq& 0,\nonumber \\
t_{i,i+{\hat{\bf y}}}(yz,\sigma) &\simeq& (-1)^{x_i
+y_i}V_{pd\pi}\theta\cos\chi.
\end{eqnarray}
where $V_{pd\pi}$ is the hopping amplitude between a Cu $d_{xy}$
orbital and an oxygen $\pi_y$ orbital.  The resulting $\vec\lambda$
vectors are then
\begin{eqnarray}
\vec\lambda_{i,i+{\hat{\bf x}}} &\simeq& {\phantom{-}} (-1)^{x_i + y_i}
(\lambda_1 \cos\chi,\lambda_2\sin\chi,0), \nonumber\\
\vec\lambda_{i,i+{\hat{\bf y}}} &\simeq& -(-1)^{x_i + y_i} (\lambda_2
\cos\chi,\lambda_1 \sin\chi,0)\label{lco}
\end{eqnarray}
and
\begin{eqnarray}
\lambda_1&\simeq& 0,\nonumber\\
\lambda_2&\simeq& {V_{pd\pi}\beta\over{\epsilon_{xz}
-\epsilon_{x^2-y^2}}}\theta,\label{lnco}
\end{eqnarray}
A non-zero $\lambda_1$, which arises to leading order from direct
hopping between neighboring $d$ orbitals, has been included in
(\ref{lco}) for completeness and also to show that such a term does
not affect the results that follow.  If we take the reasonable values
of $\beta \simeq 0.1$ eV, $\epsilon_{xz} - \epsilon_{x^2-y^2} \simeq
1.0 eV$ and $V_{pd\pi}\simeq 1.0$ eV we obtain
\begin{eqnarray}
\lambda_1 &\simeq& 0 \nonumber\\
\lambda_2 &\simeq& 100~{\rm meV}~\theta \label{lncov}
\end{eqnarray}

\subsection{YBa$_2$Cu$_3$O$_{6+x}$}

The oxygen ion displacement in a buckled CuO$_2$-Y-CuO$_2$ bilayer,
like the tilting in La$_{2-x}$Nd$_x$CuO$_4$, results in a non-zero
$t_{ij}(xz,\sigma)$ and $t_{ij}(yz,\sigma)$.  Because the geometry is
different (the $d$ orbitals are not rotated, see Fig.~\ref{fig1}b) the
proportionality to $\theta$ of these hopping integrals is larger than
in La$_{2-x}$Nd$_x$CuO$_4$ by a factor which we estimate to be roughly
4.8 \cite{harrison}, and so in this case
\begin{eqnarray}
t_{i,i+\hat{\bf x}}(xz,\sigma) &\simeq& 4.8~V_{pd\pi}\theta,\nonumber\\
t_{i,i+\hat{\bf x}}(yz,\sigma) &\simeq& 0
\end{eqnarray}
and
\begin{eqnarray}
t_{i,i+\hat{\bf y}}(xz,\sigma) &\simeq& 0,\nonumber\\
t_{i,i+\hat{\bf y}}(yz,\sigma) &\simeq& 4.8~V_{pd\pi}\theta
\label{tij}
\end{eqnarray}
and the resulting $\vec\lambda$ vectors are
\begin{eqnarray}
\vec\lambda^{(1)}_{i,i+{\hat{\bf x}}} &=&
(0,\lambda,0),\nonumber\\
\vec\lambda^{(1)}_{i,i+{\hat{\bf y}}} &=&
-(\lambda,0,0)\label{ybco2x}
\end{eqnarray}
where
\begin{equation}
\lambda\simeq{4.8~V_{pd\pi}
\beta\over{\epsilon_{xz}-\epsilon_{x^2-y^2}}}\theta\label{l123}
\end{equation}
(We will refer to the upper and lower layers as 1 and 2,
respectively.)  If the crystal field splitting in
YBa$_2$Cu$_3$O$_{6+x}$ and La$_{2-x}$Nd$_x$CuO$_4$ are assumed to be
the same then $\lambda_{123} \simeq 5\lambda_{214}$.

An important point for what follows is that because of the
inversion symmetry of the unit cell of YBa$_2$Cu$_3$O$_{6+x}$ the
$\vec\lambda_{ij}$ vectors in the lower layer are the opposite of
those in the upper layer and so
\begin{eqnarray}
\vec\lambda^{(2)}_{i,i+{\hat{\bf x}}} &=&
-(0,\lambda,0),\nonumber\\
\vec\lambda^{(2)}_{i,i+{\hat{\bf y}}} &=&
(\lambda,0,0).\label{ybco2y}
\end{eqnarray}

\section{Magnetic Anisotropy in \lnco}

In order to derive the effective spin Hamiltonian for
La$_{2-x}$Nd$_x$CuO$_4$ it is necessary to consider the large $U/t$
limit of (\ref{tb}) with $\vec\lambda_{ij}$ vectors given by
(\ref{lco}) when there is one electron per site.  Before doing this it
is useful to analyze the structure of this model by considering the
motion it describes for a single electron. For this purpose it is
natural to decompose $\vec\lambda_{ij}$ into `frustrated' and
`unfrustrated' components as follows
\begin{eqnarray}
\vec\lambda_{i,i+{\hat{\bf x}}} &=& (-1)^{x_i + y_i}
\Bigl \{\alpha_1 (\cos\chi,\sin\chi,0)+\alpha_2 (\cos\chi,-\sin\chi,0)\Bigr\}\\
 \vec\lambda_{i,i+{\hat{\bf y}}} &=& -(-1)^{x_i + y_i}
\Bigl \{\alpha_1 (\cos\chi,\sin\chi,0)-\alpha_2
(\cos\chi,-\sin\chi,0)\Bigr\}\label{decomp1}
\end{eqnarray}
where $\alpha_1 = (\lambda_1+\lambda_2)/2$ and $\alpha_2 =
(\lambda_1-\lambda_2)/2$.  To see why this decomposition is useful
consider the hopping of an electron around a single plaquette
according to (\ref{tb}) for (i) $\alpha_1 \ne 0$, $\alpha_2 = 0$, and
(ii) $\alpha_1 = 0$, $\alpha_2 \ne 0$.  For case (i) (unfrustrated)
the sign of the spin precession oscillates and no net precession
occurs (Fig.~\ref{fig2}a) while for case (ii) (frustrated) the sign of
the precession does not oscillate and there is a net precession of
$\simeq 4\alpha_1/t$ (Fig.~\ref{fig2}b).  As emphasized by SEA
\cite{sea} case (i) is special: When $\alpha_2 = 0$ it is possible to
perform a unitary transformation which maps (\ref{tb}) onto a
precession-free (isotropic) Hamiltonian.  The non-trivial physics is
thus due to the frustrated precession and it is precisely this which
is described by (\ref{pureflux}).

The unitary transformation which eliminates the unfrustrated
precession in (\ref{tb}) corresponds to a local rotation in
spin-space.  For $\theta,\alpha_2/t \ll 1$ this unitary transformation
is represented by the operator (see Appendix)
\begin{equation}
U = \exp\left\{i\sum_j {(-1)^{(x_j +y_j)} {\alpha_2\over{t}}
(\cos\chi,-\sin\chi,0)\cdot{{\vec S}_j}}\right\}\label{trans}
\end{equation}
under which (\ref{tb}) becomes, to leading order in
$\alpha/t$,
\begin{equation}
U H U^\dagger \simeq \sum_{\scriptstyle{\langle ij
\rangle}\atop{\scriptstyle\alpha\beta}}
\left\{c^\dagger_{i\alpha}
\left(-t\delta_{\alpha\beta}
+ i \vec\lambda^\prime_{ij} \cdot \vec\sigma_{\alpha\beta}
\right) c_{j\beta}+{\rm h.c.} \right\}
+U\sum_{i}
n^{\phantom{\dagger}}_{i\uparrow}n^{\phantom{\dagger}}_{i\downarrow},\label{tb2}
\end{equation}
where $\vec\lambda^\prime_{i,i+{\hat{\bf x}}} = (-1)^{x_i + y_i}
\alpha_1 (\cos\chi,\sin\chi,0)$ and
$\vec\lambda^\prime_{i,i+{\hat{\bf y}}} =
-\vec\lambda^\prime_{i,i+{\hat{\bf x}}}$.  If a global rotation in
spin space is then performed to bring the $z$-axis parallel to
$(\cos\chi,\sin\chi,0)$ the result is (\ref{pureflux}) with
$\phi_{i,i+{\hat{\bf x}}} = (-1)^{(x_i+y_i)}\tan^{-1}(\alpha_1/t)\sim
(-1)^{(x_i+y_i)}0.05\theta$ and $\phi_{i,i+{\hat{\bf y}}} =
-\phi_{i,i+{\hat{\bf x}}}$.  It follows that to leading order in
$\theta$ the energy spectrum of (\ref{tb}) is independent of the tilt
axis angle $\chi$ and depends only on the octahedral tilt angle
$\theta$.

At half-filling and when $U \gg t$ the effective spin Hamiltonian
arising from (\ref{pureflux}) is (up to an irrelevant constant)
\cite{spirals}
\begin{equation}
H_{SE} = J\sum_{\langle ij \rangle}\Bigl\{S_i^z S_j^z +
\cos\phi_{ij}(S_i^x S_j^x + S_i^y S_j^y)
+\sin\phi_{ij}(S_i^x S_j^y - S_i^y
S_j^x)\Bigr\}.\label{se214}
\end{equation}
The $\sin\phi_{ij}J{\hat{\bf z}} \cdot({\bf S}_i \times {\bf S}_j)$
term in (\ref{se214}) is minimized by a four sublattice state which
only becomes stable when the magnitude of this term is larger than $J$
\cite{cbt}.  This is never the case here and so this term is
completely frustrated, there is no spin canting, and the remaining
$\cos\phi_{ij}$ easy-axis lines the spins up parallel to the $z$
direction.

The classical ground state of (\ref{se214}) has no weak ferromagnetic
moment and so seems to disagree with experiment.  However, this is no
longer true in the physical spin-space coordinate system.  Once the
local rotations used to transform (\ref{tb}) into (\ref{pureflux}) and
also the global rotation which brought the $z$-axis in spin space
parallel to $(\cos\chi,\sin\chi,0)$ are undone the spins which once
were parallel to the $z$ axis become nearly aligned along the
$(\cos\chi,-\sin\chi,0)$ direction except for a slight cant out of the
Cu-O plane with the canting angle given by
\begin{equation}
\Theta_{wf} = {\alpha_2\sin 2\chi\over{t}}\sim0.05\theta\sin 2\chi.\label{wfm}
\end{equation}
The spin configuration in the physical basis for a specific $\chi$
between 0 and $\pi/4$ is shown in Fig.~\ref{fig3}, together with the
frustrated and unfrustrated spin precession axes.  In the LTO phase
$(\chi = \pi/4)$ (\ref{wfm}) agrees with experiment: the spins point
nearly along the $\langle 1\overline{1}0 \rangle$ direction ({\it
i.e.}, parallel to the orthorhombic {\bf c} axis) and cant out of the
plane through the angle $\Theta_{wf} \simeq 0.005$
\cite{wf}.  As $\chi$ decreases the ratio of the weak ferromagnetic
moment to $\theta$ also decreases and this decrease should be
experimentally observable in the {\it Pccn} phase of
La$_{2-x}$Nd$_x$CuO$_4$.

Given the classical ground state of (\ref{se214}) the linearized
spin-wave spectrum can be calculated using standard methods
\cite{keffer}.  The result is a twofold degenerate spectrum
defined within the AFM zone with dispersion (here $\vec q$ is in
units of one over the lattice spacing)
\begin{equation}
\omega(\vec q) = 2JS\Biggl(4-\cos^2\phi(\cos q_x + \cos q_y)^2
-\sin^2\phi(\cos q_x - \cos q_y)^2\Biggr)^{1\over 2}\label{lcosw}
\end{equation}
The two zone-center modes, which in the physical basis correspond to
in-plane and out-of-plane excitations, have a gap of $4JS\sin\phi$.
For the parameters used above this is $\simeq 1.5$ meV which agrees at
least partly with the observed zone-center spin-wave gaps of $\sim
1.0$ meV and $\sim 2.5$ meV in La$_2$CuO$_4$ \cite{swg1,swg2}.  Note,
however, that the theory presented here unambiguously predicts that
the in-plane and out-of-plane spin-wave gaps should be equal, while
experimentally these gaps have different values.  It is probable that
this discrepancy is due to dipolar interaction between spins, another
source of anisotropic spin interactions which has not been included in
this calculation.

\section{Magnetic Anisotropy in \ybcox}

Before studying the properties of a CuO$_2$-Y-CuO$_2$ bilayer it is
useful to first consider the simple toy model shown in
Fig.~\ref{fig4}: Two coupled chains in the presence of a buckling
distortion.  This model has the advantage of simplicity and the basic
physics is similar to that of the more complex two-dimensional
bilayer.  The spin Hamiltonian for this system can be written
\begin{equation}
H^{(1D)} = H^{(1D)}_1 + H^{(1D)}_2 + H^{(1D)}_{12}\label{chain}
\end{equation}
where
\begin{eqnarray}
H^{(1D)}_1 &=& J\sum_{i} \Bigl\{S^x_{1,i} S^x_{1,i+1}
+\cos\phi(S^y_{1,i} S^y_{1,i+1}+S^z_{1,i} S^z_{1,i+1})
+\sin\phi(S^y_{1,i}S^z_{1,i+1}-S^z_{1,i} S^y_{1,i+1})
\Bigr\},\nonumber\\ \label{chain1}\\
H^{(1D)}_2 &=& J\sum_{i} \Bigl\{S^x_{2,i} S^x_{2,i+1}
+\cos\phi(S^y_{2,i} S^y_{2,i+1}+S^z_{2,i} S^z_{2,i+1})
-\sin\phi(S^y_{2,i} S^z_{2,i+1}-S^z_{2,i} S^y_{2,i+1})
\Bigr\}\nonumber\\ \label{chain2}
\end{eqnarray}
and where we also include an isotropic interchain coupling
\begin{equation}
H^{(1D)}_{12} = J_{12}\sum_i S_{1,i} \cdot S_{2,i}.
\end{equation}

First consider the case $J_{12}=0$.  If we define the unitary operators
\begin{equation}
U_a(\phi) = \exp\left\{i\phi \sum_i x_i S^x_{a,i}\right\}.\label{lsm}
\end{equation}
then it is possible to transform away the DM interactions in the two
chains as follows:
\begin{eqnarray}
U_1(\phi)H^{(1D)}_1 U_1^\dagger(\phi) &=& J\sum_i\vec S_{1,i}\cdot\vec
S_{1,i+1}\label{rot1}
\\ U_2(-\phi)H^{(1D)}_2 U_2^\dagger(-\phi) &=& J\sum_i\vec S_{2,i}\cdot\vec
S_{2,i+1}.\label{rot2}
\end{eqnarray}
The eigenvalues of $H_1$ and $H_2$ are therefore the same as those of
two uncoupled isotropic Heisenberg models. The corresponding
eigenvectors are, however, different.  For example, consider the
classical ground state manifold of $H_1$.  This manifold is infinitely
degenerate because (\ref{rot1}), which is a unitary equivalent of
$H_1$, is isotropic in spin space; but the ground states of $H_1$ are
not the same as the ground states of (\ref{rot1}).  Two extreme cases
are (i) the spins lie in the $yz$-plane and form a spiral with pitch
angle $\phi$, and (ii) the spins are parallel to the $x$-axis and form
a N\'eel state.  For both cases the energy gain per site over having
the spins lying uncanted in the $yz$ plane is $JS^2\sin\phi$ and so
these states are degenerate.

The system changes qualitatively when the slightest interchain
coupling is introduced.  Figure \ref{fig4} shows one of the degenerate
classical ground states of the system when $J_{12} = 0$ in which the
spins spiral in the $yz$-plane. Note that the senses of the spirals in
the two chains are opposite.  Any finite $J_{12}$ will thus frustrate
this spiral and the spins will prefer to point in the $x$ direction so
that $J_{12}$ is unfrustrated.  The classical ground state is then
\begin{eqnarray}
\vec S_{1,i} &=& S(1,0,0)(-1)^i, \nonumber\\
\vec S_{2,i} &=& S(1,0,0)(-1)^{i+1}. \label{chaings}
\end{eqnarray}

The linearized spin-wave spectrum about (\ref{chaings}) consists of
two twofold degenerate branches with dispersions
\begin{eqnarray}
\omega^{(\pm)}(q) = 2JS\Biggl(1+2J_{12}/J-\cos^2\phi\cos^2 q
-\sin^2\phi \sin^2 q \pm 2\cos\phi\cos
q\sqrt{(J_{12}/J)^2+\sin^2\phi\sin^2q}\Biggr)^{1\over 2}.\nonumber\\
\label{swdmchain}
\end{eqnarray}
In Fig~\ref{fig5}a these two spin-wave dispersions are plotted for
small $q$, $J_{12}=0$ and $\phi = 0.02$.  As expected, in this case
the energy spectrum is the same as for two isotropic spin chains, the
only difference being that the zeros of the spectra have been shifted
from $q=0$ to $q=\pm\phi$ (this shift occurs because $U(\pm\phi)$ does
not commuting with the translation operator).  Figure \ref{fig5}b then
shows $\omega^{(-)}(q)$ for $J_{12} = 0.01 J$.  For these parameters
the spectrum has evolved into a low lying `acoustic' branch and a high
energy `optic' branch (not shown in the figure) in which the spins in
different chains precess, respectively, in and out-of relative AFM
phase.  Note that the $q=0$ acoustic mode has acquired a gap because
of the Ising-like isotropy induced by $J_{12}$.

The corresponding problem for two coupled planes is somewhat more
complex. The effective Hamiltonian for a single CuO$_2$-Y-CuO$_2$
bilayer is again of the form
\begin{equation}
H^{(2D)} = H^{(2D)}_1 + H^{(2D)}_2 + H^{(2D)}_{12}\label{planes}
\end{equation}
where $H^{(2D)}_1$ describes the `upper' layer in which the in-plane
oxygen ions are buckled downwards
\widetext
\begin{eqnarray}
H^{(2D)}_1=J\sum_i
\Bigl\{&&S^y_{1,i}S^y_{1,i+{\hat{\bf x}}}+
\cos\phi(S^x_{1,i}S^x_{1,i+{\hat{\bf x}}}+
S^z_{1,i} S^z_{1,i+{\hat{\bf x}}})+\sin\phi
(S^x_{1,i}S^z_{1,i+{\hat{\bf x}}}-S^z_{1,i} S^x_{1,i+{\hat{\bf
x}}})\nonumber\\ &&+S^x_{1,i}S^x_{1,i+{\hat{\bf
y}}}+\cos\phi(S^y_{1,i}S^y_{1,i+{\hat{\bf y}}} +S^z_{1,i}
S^z_{1,i+{\hat{\bf y}}})-\sin\phi (S^y_{1,i}S^z_{1,i+{\hat{\bf y}}}-
S^z_{1,i}S^y_{1,i+{\hat{\bf y}}})\Bigr\},\nonumber\\
\label{plane1}
\end{eqnarray}
$H^{(2D)}_2$ describes the `lower' layer in which the in-plane oxygen
ions are buckled upwards
\widetext
\begin{eqnarray}
H^{(2D)}_2=J\sum_i\Bigl\{&&S^y_{2,i}S^y_{2,i+{\hat{\bf
x}}}+\cos\phi(S^x_{2,i}S^x_{2,i+{\hat{\bf x}}} +S^z_{2,i}
S^z_{2,i+{\hat{\bf x}}})-\sin\phi (S^x_{2,i}S^z_{2,i+{\hat{\bf x}}}
-S^z_{2,i} S^x_{2,i+{\hat{\bf x}}})\nonumber\\
&&+S^x_{2,i}S^x_{2,i+{\hat{\bf
y}}}+\cos\phi(S^y_{2,i}S^y_{2,i+{\hat{\bf y}}} +S^z_{2,i}
S^z_{2,i+{\hat{\bf y}}})+\sin\phi (S^y_{2,i}S^z_{2,i+{\hat{\bf y}}}-
S^z_{2,i} S^y_{2,i+{\hat{\bf y}}})\Bigr\},\nonumber\\
\label{plane2}
\end{eqnarray}
\narrowtext\noindent
and we again assume that the layers are coupled by
\begin{equation}
H^{(2D)}_{12}=J_{12}\sum_i \vec S_{1,i} \cdot
\vec S_{2,i}.
\end{equation}

The Hamiltonians (\ref{plane1}) and (\ref{plane2}) are obtained as
before by taking the large $U/t$ limit of (\ref{tb}) with
$\vec\lambda_{ij}$ vectors given by (\ref{ybco2x}) and (\ref{ybco2y}).
Here $\phi = \lambda/t$ which we estimate to be $\sim 0.02$.  A
similar Hamiltonian (without the $\cos\phi$ terms) was derived by
Coffey, Rice and Zhang \cite{crz} for a single Cu-O layer in
YBa$_2$Cu$_3$O$_{6+x}$.  These authors noted that the classical ground
state of a single layer is not a (possibly canted) N\'eel state.
Instead, because $\phi_{ij}$ does not alternate in sign as in
La$_{2-x}$Nd$_x$CuO$_4$ there is no spin canting and the spins want to
form a spiral \cite{crz,cbt}.  It is interesting to note that the
problem of finding the classical ground state of (\ref{planes}) is
nontrivial because bonds which point in the $x$-direction favor
spiraling about the $y$-axis, while bonds which point in the
$y$-direction favor spiraling about the $x$-axis and there is no
classical ground state which spirals in this way.  Fortunately, any
difficulties associated with finding the classical ground state of
(\ref{plane1}) are probably irrelevant; just as for the coupled chains
discussed above the spiral tendencies in the two layers are opposed to
one another and so interlayer exchange, if sufficiently strong, will
lock in a commensurate AFM state.

Consider the set of classical N\'eel state parameterized as
\begin{eqnarray}
\vec
S_{1,i}&=&S(\cos\gamma\cos\eta,\sin\gamma\cos\eta,\sin\eta)(-1)^{(x_i
+ y_i)}\nonumber \\
\vec
S_{2,i}&=&S(\cos\gamma\cos\eta,\sin\gamma\cos\eta,\sin\eta)(-1)^{(x_i +
y_i+1)}.
\label{nstates}
\end{eqnarray}
Treating these as variational states for (\ref{planes}) yields an
energy per site of
\begin{equation}
E[\phi,\eta] =
-JS^2(\cos\phi(\sin^2\eta+1)+\cos^2\eta)-J_{12}S^2\label{js2}
\end{equation}
which is minimized when $\eta = 0$.  Thus the lowest energy N\'eel
states are those in which the spins lie in the $xy$ plane.  The origin
of this `easy plane' is rather subtle.  It arises from the
$\cos\phi_{ij}$ easy-axis terms in (\ref{planes}).  These terms favor
spins parallel to the $x$ direction on bonds which point in the $y$
direction, and {\it visa versa}, so that when a classical N\'eel state
is rotated in the plane the energy gain on one type of bond increases
while that on the other type decreases in such a way that the total
energy remains constant.

The classical spin-wave spectrum about (\ref{nstates}) when $\eta = 0$
can again be calculated using standard methods.  The resulting
spectrum has four branches with dispersions given by
\widetext
\begin{eqnarray}
\omega_a^{(\pm)}(\vec q) =
2JS \Biggl (&(&A(0)+2J_{12}/J)A(0)-A(q)B(\vec q)-C(\vec q)^2
\nonumber\\ &\pm&
\Bigl ( (A(0)B(\vec q)-(A(0)+2J_{12}/J)A(\vec q))^2 +4C(\vec q)^2
A(\vec q) B(\vec q)\Bigr)^{1\over{2}}\Biggr)^{1\over{2}}\nonumber\\
\\
\omega_b^{(\pm)}(\vec q) =
2JS \Biggl (&(&A(0)+2J_{12}/J)A(0)-A(\vec q)B(\vec q)-C(\vec
q)^2\nonumber\\ &\pm&
\Bigl ( ((A(0)+2J_{12}/J)B(\vec q)-A(0)A(\vec q))^2 +4C(\vec q)^2
A(\vec q) B(\vec q)\Bigr)^{1\over{2}}\Biggr)^{1\over{2}}\nonumber\\
\label{bla}
\end{eqnarray}
where
\begin{eqnarray}
A(\vec q) &=& (\cos^2\gamma + \sin^2\gamma\cos\phi)\cos q_x +
(\sin^2\gamma +
\cos^2\gamma\cos\phi)\cos q_y, \nonumber\\
B(\vec q) &=& \cos\phi(\cos q_x + \cos q_y), \nonumber\\
C(\vec q) &=& \sin\phi(\sin\gamma\sin q_x + \cos\gamma\sin q_y).
\label{blb}
\end{eqnarray}
\narrowtext
It is now apparent that the N\'eel state (\ref{nstates}) with $\eta =
0$ is not stable for small $J_{12}/J$.  In the limit $q_x, q_y, \phi
\ll 1$, $J_{12}/J\alt\phi^2$, and $\gamma = 0$ the dispersion of
the lowest lying branch is well approximated by
\begin{equation}
\omega_a^{(-)}(\vec q)
\simeq 2JS\Biggl(\left({8J_{12} -6 J \phi^2 \over{4J_{12} +
J \phi^2}}\right)q_x^2+2q_y^2\Biggr)^{1\over{2}}.\label{inst}
\end{equation}
When $J_{12}/J < 3\phi^2 /4$ the frequencies of this branch become
complex for a region in $q$-space near $\vec q = 0$.  These complex
frequencies signal the appearance of unstable modes which grow
exponentially with time.

In Fig.~\ref{fig6} the spin-wave dispersions given by (\ref{bla}) and
(\ref{blb}) are shown for $\gamma = 0$, $q_y = 0$ and $q_x$ small when
$\phi = 0.02$ and $J_{12}/J = 0.0$, $\phi^2/2$, $3\phi^2/4$ and
$0.01$.  The hatched regions denote $q$ values for which the spin wave
frequencies become complex indicating that the N\'eel state is
unstable.  As $J_{12}$ is increased the spin-wave spectrum evolves in
the following way.  First, for $J_{12} = 0.0$ there are two twofold
degenerate branches, the lower two of which become complex at small
$q_x$.  Then, as $J_{12}/J$ is gradually increased the degenerate
branches split and the region of instability in $q$-space shrinks so
that when $J_{12}/J\simeq \phi^2/2$ there is only one unstable branch
($\omega^{(-)}_a$) and when $J_{12}/J \simeq 3\phi^2/4$ the complex
frequencies disappear entirely from the spectrum and the N\'eel state
becomes locally, and almost certainly globally, stable.

For the physically relevant case $J \gg J_{12} \gg \phi^2 J$ the
dispersion of the four spin-wave branches are approximately given by
\begin{eqnarray}
\omega_a^{(-)}(\vec q) &\simeq& 2\sqrt{2}JS|\vec q| \\
\omega_b^{(-)}(\vec q) &\simeq& 2\sqrt{2}JS\sqrt{\phi^2+|\vec q|^2}\\
\omega_a^{(+)}(\vec q) &\simeq& \omega_b^{(+)}(\vec q) \simeq
2JS\sqrt{8J_{12}/J+|\vec q|^2}.\label{largej12}
\end{eqnarray}
Such a spin-wave spectrum is precisely what is expected for a bilayer
made up of two easy-plane antiferromagnets \cite{alloul}.  In this
limit $\omega_a^{(-)}(\vec q)$ and $\omega_b^{(-)}(\vec q)$ have
evolved into low energy modes corresponding to gapless in-plane and
gapped out-of-plane spin-waves, respectively.  These modes are
`acoustic' modes in the sense defined above: the spins in the two
layers precess in relative AFM phase with one another.  Figure
\ref{fig6}d shows these low lying modes for $J_{12}/J \simeq 0.01$.
The out-of-plane mode has a gap of $2\sqrt{2}JS\phi$ a result which is
consistent with the experimental observation of a gapped out-of-plane
zone-center mode \cite{rm}.  The remaining branches
$\omega_a^{(+)}(\vec q)$ and $\omega_b^{(+)}(\vec q)$ have evolved
into the high-energy `optic' modes in which spins in the two layers
precess out of AFM phase and have a gap of $4\sqrt{2}S\sqrt{J_{12}J}$.
These modes have not been seen experimentally for energies up to $\sim
50$ meV \cite{rm} indicating that $J_{12} > 0.01J$.  The interlayer
coupling is therefore large enough to lock in the N\'eel state
according to the above scenario.

\section{Conclusions}
In this paper the anisotropic corrections to superexchange arising
from SO coupling in the distorted Cu-O layers in
La$_{2-x}$Nd$_x$CuO$_4$ (tilting distortion) and
YBa$_2$Cu$_3$O$_{6+x}$ (buckling distortion) have been calculated
using Moriya's method \cite{moriya}.  Special care has been taken to
include the higher-order symmetric anisotropy terms whose importance
has recently been emphasized by SEA
\cite{sea}.

In the La$_{2-x}$Nd$_x$CuO$_4$ system in the presence of a tilting
distortion it was shown that regardless of the tilt axis it is always
possible to find a local coordinate system in spin space in which the
SO induced spin precession of moving electrons has the same form.
This form is precisely that which was studied in \cite{spirals} and is
described by Hamiltonian (\ref{pureflux}).  In this special spin space
coordinate system the classical ground state of the effective spin
Hamiltonian has no weak ferromagnetic moment but when the system is
viewed in the physical coordinate system a weak ferromagnetic moment
appears.  The ratio of the octahedral tilt angle to the weak
ferromagnetic moment depends on the tilt axis according to (\ref{wfm})
and so it decreases as the tilt axis moves from $\langle
1\overline{1}0\rangle$ in the LTO phase to $\langle 100\rangle$ in a
hypothetical insulating AFM LTT material.  The experimental
observation of the reduction of this ratio in the {\it Pccn} phase of
La$_{2-x}$Nd$_x$CuO$_4$ would provide a test of the theory discussed
here and in \cite{sea}.  One unresolved problem with the theory is
that it predicts that regardless of the tilt axis there should be only
{\it one} spin-wave gap -- a prediction which is at odds with the
experimental observation of two different gaps for the in-plane and
out-of-plane zone-center modes.  It is likely that this discrepancy is
due to the additional anisotropy caused by dipolar interactions
between the spins. However, a full treatment of the problem including
these interactions has not yet been carried out.

When the same method for calculating the SO corrections to
superexchange in La$_{2-x}$Nd$_x$CuO$_4$ is applied to the buckled
CuO$_2$-Y-CuO$_2$ bilayers in YBa$_2$Cu$_3$O$_{6+x}$ the resulting
Hamiltonian is quite different from what has been used previously to
model this system.  The Hamiltonian obtained here describes a system
in which the spins in each individual Cu-O layer tend to form a spiral
pattern.  However, one consequence of the inversion symmetry of the
unit cell of YBa$_2$Cu$_3$O$_{6+x}$ is that the senses of the spirals
favored by the two Cu-O planes in a given bilayer are opposed to one
another.  When a weak interlayer coupling is included the spiraling
become frustrated and a commensurate AFM state is stabilized.  For
physical parameters the low-energy spin-wave spectrum agrees with what
is seen experimentally: there is an in-plane gapless mode and an
out-of-plane gapped mode.

We conclude on a speculative note.  In \cite{spirals} it was argued
that the frustrated DM terms in La$_2$CuO$_4$ can give rise to new
physics away from half filling.  In particular, doped holes can gain
energy from their SO induced spin precession as they move through a
commensurate spin background.  This energy gain is similar to that
which occurs when holes move through a spiral state \cite{spirals}.
We have shown here that frustrated DM terms are present in
YBa$_2$Cu$_3$O$_{6+x}$, and the size of these terms is significantly
larger than in La$_2$CuO$_4$ (by roughly a factor of five).  Because
of this, holes doped into a CuO$_2$-Y-CuO$_2$ bilayer can also gain
energy through their spin precession.  In this case the maximum energy
gain occurs when the spins form a commensurate N\'eel background and
are lined up parallel to the $z$ direction.  When moving through such
a spin background doped holes see, effectively, a double spiral
\cite{kane}.  We note that a strong tendency for commensurate spin
fluctuations in superconducting YBa$_2$Cu$_3$O$_{6+x}$ has been
observed in both NMR \cite{nmr} and neutron scattering
\cite{rm,tranq} experiments while in doped La$_2$CuO$_4$ spin fluctuations
appear to be incommensurate \cite{aeppli,hammel}.  It is tempting to
speculate that SO effects such as those discussed here and in
\cite{spirals} may play some role in determining the different spin
dynamics of these two systems.

\acknowledgements
The author acknowledges useful discussions with A. Aharony, D. Arovas,
J. Axe, R. Hlubina, T.M. Rice, H. Vierti\"o and F.C. Zhang.  This work
was supported by a grant from the Swiss National Fund.

\widetext
\unletteredappendix{}

In this appendix the exact unitary transformation which maps
Hamiltonian (\ref{tb}) into (\ref{pureflux}) is constructed for
arbitrary $\chi$ and $\theta$.  This transformation is most easily
represented as a sequence of two local rotations in spin space.  The
first rotation eliminates the component of the unfrustrated precession
perpendicular to the frustrated axis and is generated by
\begin{equation}
U^{(1)} = \exp\left\{i\sum_j {(-1)^{(x_j +y_j)} i \tan^{-1} ({\alpha_2
\sin 2\chi\over{t}}) (\cos\chi,-\sin\chi,0)
\cdot \vec S_{j}}\right\}\label{trans1}
\end{equation}
When $U^{(1)}$ is applied to Hamiltonian (\ref{tb}) the result is
\begin{eqnarray}
U^{(1)}H{U^{(1)}}^\dagger = H^\prime =&& \sum_{\scriptstyle{\langle ij
\rangle}\atop{\scriptstyle\alpha\beta}} \left\{c^\dagger_{i\alpha}
\left(\sqrt{t^2+(\alpha_2\sin 2\chi)^2}\delta_{\alpha\beta}
+ i \vec\lambda^\prime_{ij} \cdot \vec\sigma_{\alpha\beta}\right)
c_{j\beta}+{\rm h.c.} \right\}\nonumber\\ &&+U\sum_{i}
n^{\phantom{\dagger}}_{i\uparrow}n^{\phantom{\dagger}}_{i\downarrow}
\label{tb3}
\end{eqnarray}
where the transformed $\vec\lambda^\prime$ vectors are
\begin{eqnarray}
\vec\lambda^\prime_{i,i+{\hat{\bf x}}} &=& (-1)^{x_i + y_i}
(\alpha_1 - \alpha_2 \cos 2\chi) (\cos\chi,\sin\chi,0),\nonumber\\
\vec\lambda^\prime_{i,i+{\hat{\bf y}}} &=& -(-1)^{x_i + y_i}
(\alpha_1 + \alpha_2 \cos 2\chi) (\cos\chi,\sin\chi,0).\label{decomp2}
\end{eqnarray}
The precession axes on all the links in the lattice are now parallel
to one another.  A further rotation generated by
\begin{equation}
U^{(2)} = \exp\left\{i\sum_j (-1)^{(x_j +y_j)}
\tan^{-1}\left({\alpha_2
\cos 2\chi \over {\sqrt{t^2+\alpha_2^2\sin^2 2\chi}}}\right)
(\cos\chi,\sin\chi,0)\cdot \vec S_j\right\}\label{trans2}.
\end{equation}
then ensures that the precession has the same magnitude on all the
links.  Upon applying this transformation the Hamiltonian becomes
\begin{equation}
U^{(2)}H^\prime U^{(2)^\dagger} = \sum_{\scriptstyle{\langle ij
\rangle}\atop{\scriptstyle\alpha\beta}} \left\{ c^\dagger_{i\alpha}
\left(\xi_{ij}\delta_{\alpha\beta}
+ i \vec\lambda^{\prime\prime}_{ij} \cdot \vec\sigma_{\alpha\beta}
\right) c_{j\beta}+{\rm h.c.} \right\}+U\sum_{i}
n^{\phantom{\dagger}}_{i\uparrow}n^{\phantom{\dagger}}_{j\downarrow}\label{tb4}
\end{equation}
where $\xi_{i,i+{\bf{\hat x}}} = \sqrt{t^2+\alpha_2^2 +
\alpha_1\alpha_2\sin 2\chi}$,\  $\xi_{i,i+{\bf{\hat y}}} =
\sqrt{t^2+\alpha_2^2-\alpha_1\alpha_2\sin 2\chi}$; and
\begin{eqnarray}
\vec\lambda^{\prime\prime}_{i,i+{\hat{\bf x}}} &=&
(-1)^{x_i + y_i} \alpha_1 (\cos\chi,\sin\chi,0)\\
\vec\lambda^{\prime\prime}_{i,i+{\hat{\bf y}}} &=&
-\vec\lambda^{\prime\prime}_{i,i+{\hat{\bf x}}}.
\end{eqnarray}
If the second order in $\theta$ corrections to the direct hopping
integral are ignored ($\xi_{ij} \simeq t$) and a global rotation in
spin space is performed so that the $z$-axis is parallel to
$(\cos\chi,\sin\chi,0)$ the result is Hamiltonian (\ref{pureflux})
with $\phi_{i,i+{\hat{\bf x}}} = (-1)^{(x_i+y_i)}
\tan^{-1}(\alpha_1/t)\sim (-1)^{(x_i+y_i)} 0.05\theta$
and $\phi_{i,i+{\hat{\bf y}}} = -\phi_{i,i+{\hat{\bf x}}}$.

%
\figure{
Copper $xz$ orbitals and oxygen $\sigma$ orbital in a Cu-O-Cu bond
parallel to the $x$ direction in the presence of (a) a tilting
distortion (La$_{2-x}$Nd$_x$CuO$_4$) and (b) a buckling distortion
(YBa$_2$Cu$_3$O$_{6+x}$).  In (a) the the $xz$ orbitals are tilted
with respect to the copper-oxide plane while in (b) they are not.  As
a result, the ratio of the Cu-O bond angle to $t(xz,\sigma)$ (the
hopping amplitude between the $xz$ and $\sigma$ orbitals) is larger by
a factor of $\sim 5$ in YBa$_2$Cu$_3$O$_{6+x}$ than in
La$_{2-x}$Nd$_x$CuO$_4$.\label{fig1}}

\figure{
Unfrustrated (a) and frustrated (b) spin precession about a single
plaquette in the presence of a tilting distortion.  The precession
shown in (a) is unfrustrated because the electron spin returns to its
original value upon hopping around any closed loop.  The precession
shown in (b) is frustrated because an electron which hops around a
single plaquette acquires a finite spin precession.  While it is
always possible to eliminate the unfrustrated precession by a unitary
transformation the frustrated precession cannot be so removed and thus
is responsible for any physical anisotropy in spin space due to
spin-orbit coupling.\label{fig2}}

\figure{
Two neighboring copper-oxide clusters in the {\it Pccn} phase of
La$_{2-x}$Nd$_x$CuO$_4$.  The heavy line is the
$(\cos\chi,\sin\chi,0)$ axis about which the CuO$_6$ octahedra are
tilted through an angle $\theta$. The light line is the
$(\cos\chi,-\sin\chi,0)$ axis.  The spin precession of a single
electron which hops in the presence of this tilting distortion is a
combination of frustrated precession about the light axis and
unfrustrated precession about the heavy axis.  The classical ground
state of the effective spin Hamiltonian describing this system at half
filling is also shown.  The spins are nearly lined up along the
frustrated axis except for a slight cant out of the plane by an angle
which is proportional to $\sin(2\chi)$.\label{fig3}}

\figure{
Two buckled CuO `chains' representing the toy model described by
(\ref{chain}).  The spin configuration shown is one possible classical
ground state in the absence of interchain coupling.  The DM
interactions induced by the buckling cause the spins to form a spiral,
but because the buckling in the bottom chain is opposite to that of
the top chain the spirals in the two chains have opposite
senses.\label{fig4}}

\figure{
Spin-wave spectrum for two buckled chains when (a) $J_{12} = 0$ and
(b) $J_{12} = 0.01 J$.  In (a) the spin wave spectrum is identical to
that of two isotropic Heisenberg models except that $q$ has been
shifted by $\pm \phi$.  In (b) there is a finite interlayer coupling
and the system has acquired an Ising-like character and a
gap.\label{fig5}}

\figure{
Spin-wave spectrum about a N\'eel state with spins parallel to the $x$
direction for a CuO$_2$-Y-CuO$_2$ bilayer in YBa$_2$Cu$_3$O$_{6+x}$
for $J_{12}/J$ equal to (a) $0$; (b) $0.25\phi$; (c) $0.75\phi$; and
(d) $0.01$ where $\phi \simeq 0.02$.  The hatched regions represent
$q$ values where one or more of the frequencies are complex indicating
an instability.  There are four modes which degenerate into two when
$J_{12}/J = 0$ (a).  As $J_{12}/J$ increases these modes split.  In
(b) one of the two unstable modes has become stable.  In (c) both have
become stable and in (d) the two higher lying `optic' modes are not
shown and the two lowest modes correspond to a gapless in-plane and a
gapped out-of-plane mode in agreement with experiment.\label{fig6}}

\begin{references}

\bibitem{moriya} T. Moriya, Phys. Rev. {\bf 120}, 91 (1960).

\bibitem{pwa} P.W. Anderson, Phys. Rev. {\bf 115}, 2 (1959)

\bibitem{dizzie} I. Dzyaloshinski, J. Phys. Chem. Solids {\bf 4}, 241
(1958).

\bibitem{note1} In this paper all corrections to superexchange due
to SO coupling will be referred to as DM interactions.  This includes
(using the notation of \cite{moriya}) antisymmetric terms of the form
$\bf D\cdot(\bf S\times\bf S)$ which are linear in SO coupling, and
symmetric terms of the form $\bf S\cdot\bf\Gamma\cdot\bf S$ which are
quadratic.

\bibitem{wf} T. Thio, T.R. Thurston, N.W. Preyer, P.J. Picone, M.A.
Kastner, H.P. Jenseen, D.R. Gabbe, C.Y. Chen, R.J. Birgeneau, and A.
Aharony, Phys. Rev. B {\bf 38}, 905 (1988); and S-W. Cheong, J.D.
Thompson, and Z. Fisk, Physica C {\bf 158}, 109 (1989).

\bibitem{rm} J. Rossat-Mignod {\it et al.}, Physica B {\bf 169}, 58
(1991), and references therein.

\bibitem{spirals} N.E. Bonesteel, T.M. Rice and F.C. Zhang, Phys. Rev.
Lett. {\bf 68} 2684, (1992).

\bibitem{sea} L. Shekhtman, O. Entin-Wohlman and A. Aharony, Phys.
Rev. Lett. {\bf 69} 836, (1992); and L. Shekhtman, A. Aharony and O.
Entin-Wohlman, preprint.

\bibitem{pccn1} M. Crawford, R.L. Harlow, E.M. McCarron, W.E. Fameth,
J.D. Axe, H. Chou, and Q. Huang, Phys. Rev. B {\bf 44}, 7749 (1991).

\bibitem{pccn2} B. B\"uchner {\it et al.}, preprint.

\bibitem{crz} D. Coffey, T.M. Rice and F.C. Zhang, Phys. Rev. B {\bf
44}, 10112 (1991).

\bibitem{bts} J.F. Bringley, S.S. Trail and B.A. Scott, Journ. of
Sol. State Chem. {\bf 86}, 310 (1990).

\bibitem{ltt} J.D. Axe, A.H. Moudden, D. Hohlwein, D.E. Cox, K.
Mohanty, A.R. Moodenbaugh, and Y., Phys. Rev. Lett. {\bf 62}, 2751
(1989).

\bibitem{chakra} S. Chakravarty, in {\it High-Temperature
Superconductivity}, edited by K.S. Bedell {\it et al.}
(Addison-Wesley, Reading, MA, 1990) and E. Manousakis, Rev. Mod. Phys.
{\bf 63}, 1 (1991) and references therein.

\bibitem{cbt} D. Coffey, K. Bedell and S.A. Trugman, Phys. Rev. B {\bf
42}, 6509 (1990).

\bibitem{swg1} C.J. Peters {\it et al.}, Phys. Rev. B {\bf 37}, 9761
(1988).

\bibitem{swg2} R.T. Collins {\it et al.}, Phys. Rev. B {\bf 37} (1988).

\bibitem{kom} W. Koshibae, Y. Ohta, and S. Maekawa, Preprint.

\bibitem{swst} J.S. Swinnea and H. Steinfink, J. Matter. Res. {\bf
2}, 424 (1987).

\bibitem{alloul} See, for example, H. Alloul, in {\it High Temperature
Superconductivity: Proceedings of the Thirty-Ninth Scottish
Universities Summer School in Physics}, edited by D.P. Tunstall and W.
Barford (Adam Hilgar, Bristol, 1992), and references therein.

\bibitem{tranq} J.M. Tranquada, P.M. Gehring, G. Shirane, S. Shamoto
and M. Sato, preprint.

\bibitem{emery} V.J. Emery and G. Reiter, Phys. Rev. B {\bf 38},
4547 (1988).

\bibitem{hyb} M.S. Hybertson, M. Schl\"uter and N.E. Christensen,
Phys. Rev. B {\bf 39}, 13 (1989).

\bibitem{slater} See for example, J.C. Slater, {\it Quantum Theory
of Matter} (McGraw-Hill, New York, 1968) pp.~306-309

\bibitem{harrison} See, for example, W. A. Harrison, {\it Electronic
Structure and the Properties of Solids} (W.H. Freeman, San Francisco,
1980).

\bibitem{keffer} F. Keffer, {\it Handbuch der Physik}, edited by S.
Fl\"ugge (Springer-Verlag, Berlin, 1966), Vol. 18, Pt. 2, p.1.

\bibitem{kane} C.L. Kane, P.A. Lee, T.K. Ng, B. Chakraborty, and N.
Read, Phys. Rev. B {\bf 41}, 2653 (1990).

\bibitem{nmr} R.E. Walstedt and J.W.W. Warren, Science {\bf 248}, 1082
(1990) and references therein.

\bibitem{aeppli} S-W. Cheong {\it et al.}, Phys. Rev. Lett. {\bf 67},
1791 (1992); and T.E. Mason, G. Aeppli, and H.A. Mook, Phys. Rev.
Lett. {\bf 68}, 1414 (1992).

\bibitem{hammel} P.C. Hammel, E.T. Ahrens, A.P. Reyes, P.C. Canfield,
Z. Fisk, J.D. Thompson and J.E. Schirber, preprint.

\end{references}
\end{document}